\documentstyle[12pt]{article}
\begin{document}
\newcommand{\bbox}{\stackrel{-}{\Box}}
\newcommand{\bnabla}{\bar {\nabla}}
\begin{center}
QUANTUM  COSMOLOGY  FROM  $ \cal {N}$ =  4    SUPER YANG - MILLS  THEORY\\
\bigskip
I. Brevik\footnote{E-mail: iver.h.brevik@mtf.ntnu.no}\\
Division of Applied Mechanics\\
Norwegian University of Science and Technology\\
N-7034 Trondheim, Norway\\
\bigskip
and\\
\bigskip
S. D. Odintsov\footnote{E-mail: odintsov@mail.tomsknet.ru}\\
Tomsk State Pedagogical University\\
634041 Tomsk, Russia\\
\end{center}
\bigskip
\begin{center}
\bigskip
February 1999\\
\end{center}
\begin{abstract}
We consider quantum $\cal{N}$  = 4  super Yang-Mills theory interacting in a covariant way with $\cal{N}$= 4  conformal supergravity. The induced large $N$ effective action for such a theory is calculated on a dilaton-gravitational background using the conformal anomaly found via  AdS/CFT correspondence. Considering such an effective action as a quantum correction to the classical gravity action we study quantum cosmology. In particular, the effect from dilaton to the scale factor (which without dilaton corresponds to the inflationary universe) is investigated. It is shown that, dependent on the initial conditions for the dilaton, the dilaton may slow down, or accelerate, the inflation process.  At late times, the dilaton is decaying exponentially. 
\end{abstract}
\newpage
 The Inflationary Universe (for a general review, see \cite{KT}) is considered to be a quite realistic element of the very early Universe evolution. Recently there appeared evidence, however, that the present Universe is  subject to accelerated expansion and may thus be entering an inflationary phase now. This calls for a reconsideration of quantum cosmology and construction of  new (or modified) versions of the theory of the very early Universe.

In the present letter, aimed towards this purpose, we consider quantum cosmology as induced from $\cal {N}$ =  4  quantum super YM theory on the background  $\cal {N}$ = 4  conformal supergravity. Such theories have become very popular recently  in connection with AdS/CFT correspondence. Using conformal anomaly on a dilaton-gravitational background we construct the anomaly induced effective action and study the consequences it may lead to in the early Universe. It is known that on a purely gravitational background such an effective action leads to the possibility of inflation. As we will show, the role of the dilaton is to accelerate, or to slow down, the inflationary expansion, depending on the choice of initial conditions for the dilaton.

 Let us start from the Lagrangian of local superconformally invariant  $\cal {N}$ = 4  super YM  theory in the background of $\cal{N}$ = 4  conformal supergravity (see \cite{FT} for introduction and review). The corresponding vector multiplet is  $(A_\mu, \psi_i, X_{ij})$. Supposing that super  YM  theory interacts with conformal supergravity in a  SU(1,1)  covariant way and keeping only kinetic terms, we get:
\begin{eqnarray}
L_{SYM}&=&-\frac{1}{4}(e^{-\phi}F_{\mu\nu}F^{\mu\nu}+C F^{\mu\nu}F_{\mu\nu}^*)- \nonumber  \\
       & &-\frac{1}{2}\bar{\psi}^i \gamma^\mu D_\mu \psi^i-
\frac{1}{4}X_{ij}(-D^2+\frac{1}{6}R)X^{ij}+...
\end{eqnarray} 
\label{1} 
Note that the complex scalar $\phi$ from the conformal supergravity multiplet is written as  $C+ie^{-\phi}$. It is also to be noted that the first term in (1) describes the dilaton coupled electromagnetic field whose conformal anomaly has been found in \cite{NO}.

Despite the fact that  $\cal{N}$ = 4  super  YM  theory is finite in Minkowski space, such a theory on a curved background contains divergences (vacuum polarization, or the action for external fields; see \cite{BOS} for a review). The standard calculation based upon the background field method  (see \cite{BOS} for an introduction) gives the vacuum polarization (divergent part of the effective action) which is the action of the  $\cal{N}$ = 4  conformal supergravity \cite{LT}. At the same time, $\Gamma_{div}$ is actually proportional to the $a_2$-coefficient of the proper time expansion. This coefficient defines also the conformal anomaly of super  YM  theory.    

On a purely bosonic background with only non-zero gravitational and dilaton fields, the conformal anomaly for ${\cal N}$ = 4  super YM theory has been calculated in \cite{NO1} via AdS/CFT correspondence \cite{MA}, \cite{WI}, \cite{GKP} to be
\begin{eqnarray}
T=b(F+\frac{2}{3}\Box R)+b'G+b''\Box R \nonumber \\
+C[\Box \phi^*\Box \phi-2(R^{\mu\nu}-\frac{1}{3}g^{\mu\nu} R)\nabla_{\mu}\phi^*\nabla_{\nu}\phi)].
\end{eqnarray}
\label{2}
Here
\[ b=\frac{N^2-1}{(4\pi)^2}\frac{N_s+6N_f+12N_v}{120}=\frac{N^2-1}{4(4\pi)^2}, \]
\[b'=-\frac{N^2-1}{(4\pi)^2}\frac{N_s+11N_f+62N_v}{360}=-\frac{N^2-1}{4(4\pi)^2}, \]
\[ C=\frac{N^2-1}{(4\pi)^2}N_v=\frac{N^2-1}{(4\pi)^2}. \]

We have used that  $N_s=6$, $N_f=2$, $N_v=1$ in ${\cal N}$ = 4  SU(N) super YM theory; \( F=R_{\mu\nu\alpha\beta}R^{\mu\nu\alpha\beta}-2R_{\mu\nu}R^{\mu\nu}+\frac{1}{3}R^2$ is the square of the Weyl tensor in four dimensions; $G$ is the Gauss-Bonnet invariant. The prefactor $N^2-1$ appears because all fields are in the adjoint representation. As the AdS/CFT correspondence gives only the large $N $ limit, only the part proportional to $N^2$ in (2) has been found in \cite{NO1} (with  accuracy up to the $\Box R$ term). Note that the coefficient $b''$ is known to be in general ambiguous. In the above super YM theory $b''$ may be equal to zero, as one can see from a direct calculation (see \cite{MD} for a review). On the other hand, in the scheme advocated in \cite{LT},  $b''$ is found to be such that it cancels out  the $\Box R $ contribution in the first term and no $\Box R $-term remains in the total conformal anomaly (2). In this last case, the conformal anomaly (including all terms, not only metric and dilaton) coincides with the action of $\cal {N}$ = 4 conformal supergravity.

It is of interest to note that the dilaton dependent term in (2) may be rewritten in the following form (up to a total derivative)
\begin{eqnarray}
\Box \phi^* \Box \phi-2(R^{\mu\nu}-\frac{1}{3}g^{\mu\nu}R)\nabla_{\mu}\phi^*\nabla_{\nu}\phi= \nonumber \\
=\phi^*[\Box^2+2R^{\mu\nu}\nabla_\mu \nabla_\nu -\frac{2}{3}R \Box +\frac{1}{3}(\nabla^\mu R)\nabla_\mu]\phi.
\end{eqnarray}
\label{3}
The r.h. side of (3) represents a conformally invariant, fourth order operator acting on scalars.

Let us now find the anomaly induced effective action \cite{RFT}. We will write it in non-covariant, local form:
\begin{eqnarray}
W	&  =	&  b \int d^4 x \sqrt{-\bar{g}} \bar{F} \sigma+ 
   b'\int d^4 x \sqrt{-\bar{g}} [ \, \sigma [2\,{\bbox}^2 \nonumber \\
	&  + 	&     4 \bar{R}^{\mu \nu} {\bnabla}_\mu {\bnabla}_\nu 
   -\frac{4}{3}\bar{R} \bbox  +\frac{2}{3}({\bnabla}^\mu \bar{R}){\bnabla}_\mu ]\sigma 
	 +(\bar{G}- \frac{2}{3} {\bbox} \bar{R})\sigma ]  \nonumber \\
	&	& -\frac{1}{12}[b''+\frac{2}{3}(b+b')] \int d^4 x \sqrt{-\bar{g}}[ \,
\bar{R} -6 \bbox \sigma -6( {\bnabla}_\mu \sigma )({\bnabla}^\mu \sigma) ]^2 \nonumber \\
	&	& +C \int d^4 x \sqrt{-\bar{g}}\, \sigma \phi^* ( {\bbox}^2 +2 {\bar{R}}^{\mu \nu}
{\bnabla}_\mu {\bnabla}_\nu-\frac{2}{3} \bar{R} \bbox +\frac{1}{3}({\bnabla}^\mu \bar{R} ){\bnabla}_\mu)\phi.
\end{eqnarray} 
\label{4} 
Note that in the conformal anomaly (2) we used $ g_{\mu\nu} = e^{2 \sigma} \bar{g}_{\mu \nu} $, and all quantities in (4) are calculated with the help of the overbar metric.

As is known, the anomaly induced effective action is defined with accuracy up to a conformally invariant functional. We limit ourselves to a conformally flat metric, i. e. $\bar{g}_{\mu \nu}= {\eta}_{\mu \nu} $. In this case, the conformally invariant functional on a purely gravitational background is zero, and  $ W $ in Eq.(4) gives the complete contribution to the one-loop effective action. (This conformally invariant functional is not zero in the presence of the dilaton. There is no way to calculate it in closed form; one has to use the Schwinger-De Witt expansion. Nevertheless, one can show that its effect on $ W $ will be small.)  We also assume that  only the real part of the dilaton is non-zero.

The anomaly induced effective action (4) may now be simplified significantly (due to the fact that $ {\bar{g}}_{\mu \nu} = {\eta}_{\mu \nu} $):
\begin{eqnarray}
W	&  =	& \int d^4 x [ \, 2 b' \sigma {\Box}^2 \sigma -3 ( b'' +\frac{2}{3}(b+b')) \times \nonumber \\
	&	& \times (\Box \sigma + \partial_\mu \sigma \partial ^\mu \sigma)^2 
+ C \sigma \,\phi \, {\Box}^2 \phi ] ,
\end{eqnarray}
\label{5}
where all derivatives are now flat ones.

We consider the case when the scale factor $ a(\eta) $ depends only on conformal time: $\sigma(\eta)=\ln a(\eta) $. Then one has to add the anomaly induced effective action to the classical gravitational action:
\begin{equation}
S_{cl} =- \frac{1}{\kappa} \int d^4 x \sqrt{-g}\, R = -\frac{6}{\kappa}\int d^4 x 
e^{2 \sigma} \, (\partial_\mu \, \sigma)^2, 
\end{equation}
\label{6}
where $ \kappa = 16 \pi G $.

Now one can write the equations of motion for the action $ S_{cl}+W $. We assume that $\sigma $ and $ \phi $ depend only on the conformal time $\eta$. The equations of motion for $S_{total}=S_{cl}+W$ have the following form:
\begin{eqnarray}
\frac{a''''}{a}& -	& \frac{4 a'\,a'''}{a^2}-\frac{3 a''^2}{a^2}+\frac{a''\,a'^2}{a^3} \left( 6-
\frac{12 b'}{3b''+2 b} \right) \nonumber \\
		  &   & + \frac{12 b'\,a'^4}{(3 b''+2 b)a^4}-\frac{6}{\kappa (3b''+2b)} a\,a''
-\frac{C}{2(3b''+2b)}\phi\,\phi''''=0, \nonumber \\
              &   & \nonumber \\ 
              &   & \ln a\; \phi'''' + (\ln a\; \phi)'''' =0,
\end{eqnarray}  
\label{7}
where $3 b''+2 b \neq 0$. As we said above, the natural choice for $b''$ is to take $b''=0$, corresponding to not making any finite renormalization of the gravitational action.

Note that when $b''=-\frac{2}{3}b $ ( corresponding to the scheme advocated in \cite{LT}), the first of Eqs.(7) simplifies:
\begin{equation}
-12 b'\frac{a''a'^2}{a^3}+12 b'\; \frac{a'^4}{a^4}-\frac{6}{\kappa}a\;a''-\frac{C}{2}\phi\;\phi''''=0,
\end{equation}
\label{8}
while the second of Eqs.(7) has the same form.

First of all, we make the transformation to cosmological time in the above equations: $ dt = a(\eta)\,d\eta $. Then the first of the equations of motion (7) takes the following form:
\begin{eqnarray}
a^2 \stackrel{....}{a}&  +&3 a \dot{a}\stackrel{...}{a}+a \ddot{a}^2-\left( 5+\frac{12 b'}{3 b''+2b} \right)
\dot{a}^2\,\ddot{a} \nonumber \\
                      &  -&\frac{6}{\kappa (3 b''+2b)}(a^2 \ddot{a}+a \dot{a}^2)-
\frac{C \phi Y[\phi,a]}{2(3b''+2b)}=0.
\end{eqnarray}
\label{9}
Here,
\begin{eqnarray*}
Y[\phi,a]&=	& a^3 \stackrel{....}{\phi}+6a^2 \dot{a}\stackrel{...}{\phi} \\
         &+	& 4a^2\ddot{a}\ddot{\phi}+7a \dot{a}^2\ddot{\phi}+4a \dot{a}\ddot{a}\dot{\phi}
+a^2\stackrel{...}{a}\dot{\phi} +\dot{a}^3 \dot{\phi}.
\end{eqnarray*} 
The second of Eqs.(7) becomes:
\begin{eqnarray}
&	& 2 a \ln a \;Y[\phi,a]+\phi a^3\stackrel{....}{a}+4a^3\dot{a}\stackrel{...}{\phi} \nonumber \\
&  +	& 3a^2\phi\,\dot{a}\stackrel{...}{a}+4a^3\dot{\phi}
\stackrel{...}{a}+6a^3\ddot{\phi}\ddot{a}+12a^2\ddot{\phi}\;\dot{a}^2 \nonumber \\
&  +	& 14a^2\dot{\phi}\dot{a}\ddot{a}+a\phi\ddot{a}\dot{a}^2+a^2\phi\ddot{a}^2
+4a\dot{\phi}\dot{a}^3=0.
\end{eqnarray}
\label{10}
As one can see, these equations are too complicated to be solved analytically in general.

  Let us consider first the solution when dilaton is absent. For the first choice $b''=0$ we get the following special solution: $a(t)=a_0\,e^{Ht} $, where $H^2=-\frac{1}{\kappa b'}$,  $a_0$ being  an arbitrary constant. This inflationary solution
(for positive $H$) has been first found by Starobinsky and Mamaev-Mostepanenko in \cite{Sta} using the renormalized EMT of conformal matter on the right-hand side of Einstein's equations. It has also been recovered in the third reference in \cite{RFT}, using the effective action approach \cite{BOS}.

For the second choice  $ b''=-\frac{2}{3}b $ we get just the the same solution $a(t)=a_0e^{Ht}$. That indicates a physical equivalence of the above two forms for the conformal anomaly (at least in some specific situations).

Now we consider the solution with non-zero dilaton. We look at the approximate special solution of Eq.(10) when the term with $\ln a$ may be dropped. Since $a(t)=a_0e^{Ht}$,  $\ln a \sim Ht$.Moreover, $H$ is proportional to the Planck mass, so $Ht$ is an extremely small quantity and our approximation is justified.

Then, we search for special solutions of the sort:
\begin{equation}
a(t) \simeq \tilde{a}_0 e^{\tilde{H}t},~~~~\phi(t) \simeq \phi_0 e^{-\alpha \tilde{H}t}
\end{equation}
\label{11}
From Eq.(10) (without logarithmic term) we obtain
\begin{equation}
\phi(t)=\phi_1 e^{-\frac{3}{2}\tilde{H}t}+\phi_2 e^{-2.62 \tilde{H}t}+
\phi_3 e^{-0.38 \tilde{H}t},
\end{equation}
\label{12}
where the $\phi_1,\, \phi_2,\,\phi_3$ are constants. Taking the particular solution $ \phi(t)=\phi_0 e^{-\alpha \tilde{H}t} $ from Eq.(12) and substituting it back in Eq.(9) we obtain
\begin{equation}
\tilde{H}^2 \simeq -\frac{1}{\kappa} \left[ b'+\frac{C}{24}\phi_0^2(\alpha^4-6\alpha^3+11\alpha^2
-6 \alpha) \right] ^{-1}.
\end{equation}
\label{13}
This solution is the same for any choice of $b''$.

As is easily seen, for $\alpha \simeq 2.62 $ or $ \alpha \simeq 0.38 $ the denominator in Eq.(13) is increasing in absolute value, being negative all the time. As a result, for this case $ \tilde{H} < H = \sqrt{-\frac {1}{\kappa b'}}$. Hence, the effect of dilaton is to slow down the inflation.

For the mode $\alpha = \frac{3}{2}$, the second term in the denominator is positive, hence $ \tilde{H}^2 > H^2$. The role of the dilaton in this case is to make the inflation faster, as compared with the case of no dilaton. Without the influence from dilaton, the duration of inflation is high enough:
\begin{equation}
\frac{\Delta t}{\sqrt \kappa} \simeq \frac{N}{4 \pi} n,
\end{equation}
\label{14}
where $ n $ is the number corresponding to expansion of the universe at $ 10^n $ times. Taking, for example, $N$ = 4 (SU(4) super YM) and taking $n \simeq 30\pi $, we obtain the necessary duration of inflation which is just two orders of magnitude greater than the Planck time. Then, for initial conditions $\phi_1 \simeq \phi_0, ~~ \phi_2 = \phi_3 = 0 $, due to faster inflation $\Delta t $ is getting smaller and the above choice of $n$ is not enough to meet the necessary requirement of duration of inflation. If $\phi_1=0$ with 
the same choice for time interval we get higher than necessary duration of inflation 
as dilaton effect.

	In summary, we have discussed quantum cosmology from $\cal{N} $ = 4 super YM theory on a dilaton-gravitational background. As one solution we found the inflationary (conformally flat) Universe with exponentially decaying dilaton. It is not difficult to consider other cosmogical models (say closed or open FRW), or other types of behaviour of the scale factor. However in this case one should do a numerical study of the effective equations of motion. The results, of course, should depend very much on the choice of initial conditions for $a(t), ~\phi(t)$ and their derivatives.

{\bf Acknowledgments} SDO thanks P. van Nieuwenhuizen and E. Mottola for useful discussions. The work of SDO has been partially supported by NATO Science Fellowship, grant 128058/410 (Norway).


\begin{thebibliography}{99}
\bibitem{KT}
 E. Kolb and M. Turner, The Very Early Universe,
Addison-Wesley,NY, 1994
\bibitem{FT}
M. Kaku, P.K. Townsend and P. van Nieuwenhuizen,
Phys. Rev. D {\bf 17}, 3179 (1978);
E. Bergshoeff, M. de Roo and B. de Wit,
Nucl. Phys. B {\bf 182}, 173 (1981);
E.S. Fradkin and A. Tseytlin,
Phys. Repts. {119}, 233 (1985).
\bibitem{NO}
S. Nojiri and S.D. Odintsov,
Phys. Lett. B {\bf 426}, 29 (1998),hep-th 9801052;
S. Ichinose and S.D. Odintsov,
Nucl. Phys. B {\bf 539}, 643 (1999),hep-th 9802043.
\bibitem{BOS}
I.L. Buchbinder, S.D. Odintsov and I.L. Shapiro,
Effective Action in Quantum Gravity,
IOP Publishing, Bristol and Philadelphia, 1992.
\bibitem{LT}
H. Liu and A. Tseytlin,
Nucl. Phys. B {\bf 533}, 88 (1998), hep-th 9804083.
\bibitem{MA}
J.M. Maldacena,
Adv. Theor. Math. Phys. {\bf 2}, 231 (1998), hep-th 9711200.
\bibitem{WI}
E. Witten,
Adv. Theor. Math. Phys. {\bf 2}, 505 (1998), hep-th 9802150.
\bibitem{GKP}
S. Gubser, I. Klebanov and A. Polyakov,
Phys. Lett. B {\bf 428}, 105 (1998), hep-th 9802109.
\bibitem{NO1}
S. Nojiri and S.D. Odintsov,
Phys. Lett. B {\bf 444}, 92 (1998),
hep-th 9810008
\bibitem{MD}
M.J. Duff,
Class. Quant. Grav. {\bf 11}, 1387 (1994);
Nucl. Phys. B {\bf 125}, 334 (1977).
\bibitem{RFT}
R. Reigert,
Phys. Lett. B {\bf 134}, 56 (1984);
E.S. Fradkin and A. Tseytlin,
Phys. Lett. B {\bf 134}, 187 (1984);
I.L. Buchbinder, S.D. Odintsov and I.L. Shapiro,
Phys. Lett. B {\bf 162}, 92 (1985);
I. Antoniadis and E. Mottola,
Phys. Rev. D {\bf 45},2013 (1992);
S.D. Odintsov,
Z. Phys. C {\bf 54}, 531 (1992).
\bibitem{Sta}
A. Starobinsky,
Phys. Lett. B {\bf 91}, 99(1980);
S.G. Mamaev and V.M. Mostepanenko,
JETP {\bf 51}, 9 (1980).
\end{thebibliography}
\end{document}